\documentclass[12pt, 
]{article}

\usepackage{graphicx}
\usepackage{epsfig}
\usepackage{wrapfig}
\usepackage{amsfonts}
\usepackage{amsmath}

\usepackage[usenames]{color}
\usepackage{amssymb}
\usepackage[mathscr]{eucal} 
\usepackage{longtable}

\usepackage{color}
\input{colordvi.tex}

\def\kesu#1{}

\begin{document}
\baselineskip 0.6cm
\newcommand{\gsim}{ \mathop{}_{\textstyle \sim}^{\textstyle >} }
\newcommand{\lsim}{ \mathop{}_{\textstyle \sim}^{\textstyle <} }
\newcommand{\vev}[1]{ \left\langle {#1} \right\rangle }
\newcommand{\bra}[1]{ \langle {#1} | }
\newcommand{\ket}[1]{ | {#1} \rangle }
\newcommand{\Dsl}{\mbox{\ooalign{\hfil/\hfil\crcr$D$}}}
\newcommand{\nequiv}{\mbox{\ooalign{\hfil/\hfil\crcr$\equiv$}}}
\newcommand{\nsupset}{\mbox{\ooalign{\hfil/\hfil\crcr$\supset$}}}
\newcommand{\nni}{\mbox{\ooalign{\hfil/\hfil\crcr$\ni$}}}
\newcommand{\EV}{ {\rm eV} }
\newcommand{\KEV}{ {\rm keV} }
\newcommand{\MEV}{ {\rm MeV} }
\newcommand{\GEV}{\mbox{ GeV}}
\newcommand{\TEV}{\mbox{ TeV}}

\def\diag{\mathop{\rm diag}\nolimits}
\def\tr{\mathop{\rm tr}}

\def\Spin{\mathop{\rm Spin}}
\def\SO{\mathop{\rm SO}}
\def\O{\mathop{\rm O}}
\def\SU{\mathop{\rm SU}}
\def\U{\mathop{\rm U}}
\def\Sp{\mathop{\rm Sp}}
\def\SL{\mathop{\rm SL}}

\def\change#1#2{{\color{blue}#1}{\color{red} [#2]}\color{black}\hbox{}}

\newcommand{\pt}{$p_T$}
\newcommand{\mt}{$M_T$}
\newcommand{\sqs}{\sqrt{s}}
\newcommand{\sqshat}{\sqrt{\hat{s}}}
\newcommand{\mtil}{\tilde{m}}
\newcommand{\mttwo}{M_{T2}}
\newcommand{\mttm}{M_{T2}^{\max}}
\newcommand{\mch}{m_{c}}
\newcommand{\mcc}{\tilde{m_{c}}}
\newcommand{\ptvec}[1]{\mathbf{p}_T^{#1} }
\newcommand{\Ptvec}[1]{\mathbf{P}_T^{#1} }
\newcommand{\veczero}{\mathbf{0} }
\newcommand{\mT}[1]{m_{T}^{(#1)} }
\newcommand{\mttmc}{M_{T2}^{max}(\tilde{m}_c, P_T)}
\newcommand{\ptvecp}[1]{\mathbf{p}_{T\|}^{#1} }
\newcommand{\Ptvecp}[1]{\mathbf{P}_{T\|}^{#1} }
\newcommand{\mttp}{M_{T2\|}}
\newcommand{\ptveco}[1]{\mathbf{p}_{T\bot}^{#1} }
\newcommand{\Ptveco}[1]{\mathbf{P}_{T\bot}^{#1} }
\newcommand{\mtto}{M_{T2\bot}}
\newcommand{\ptmiss}{\ptvec{miss}}
\newcommand{\Etmiss}{E_T^{miss}}
\newcommand{\ttbar}{t\bar{t}}
\newcommand{\bbbar}{b\bar{b}}
\newcommand{\lag}{\mathcal{L}} 
\newcommand{\mtop}{m_{\rm top}}
\newcommand{\MG}{{\sc  \sc MadGraph/MadEvent 4.4} }
\newcommand{\PY}{{\sc Pythia}}
\newcommand{\fb}{fb$^{-1}$ }


\begin{titlepage}
  
\begin{flushright}
IPMU11-0104\\
\end{flushright}
   
\vskip 1cm
\begin{center}
  
 {\large \bf Improved discovery of a nearly degenerate model: \\MUED using $M_{T2}$ at the LHC}

\vskip 1.2cm
   
{ Hitoshi Murayama$^{a,b}$, Mihoko M. Nojiri$^{b,c}$, and Kohsaku Tobioka$^{b,d}$}
    
\vskip 0.4cm
 
{ \it
  $^a$Department of Physics, University of California, Berkley CA94720\\
  Theoretical Physics Group, Lawrence Berkley National Laboratory,
 Berkley CA94720
   \\[1.5mm]
          
  $^b$Institute for the Physics and Mathematics of the Universe, \\University of Tokyo, Kashiwa  277-8583, Japan
\\[1.5mm]
  $^c$Theory Group, KEK, Tsukuba 305-0801, Japan \\The Graduate University for Advanced Studies (Sokendai), Tsukuba 305-0801, Japan
   \\[1.5mm]
 $^d$Department of Physics, University of Tokyo, Tokyo 113-0033, Japan 
   \\[1.5mm]             
}
\vskip 1.5cm
   
\abstract{We study the discovery potential of the Minimal Universal
 Extra Dimension model (MUED) and improve it utilizing the multijet + lepton
 mode at the LHC.
Since the MUED has a nearly degenerate spectrum, most events only have soft jets and small
 $E_T^{miss}$. The signature is challenging to
 search. 
We apply $M_{T2}$ for the event selection and set the invisible
 particle mass of $M_{T2}$ (test mass) to zero. The test mass is much smaller than
 the invisible particle mass of MUED. In that case, 
$M_{T2}$ of the
 signal can be large depending on Up-Stream Radiations (USR) which includes initial state
 radiations (ISR). 
  On the other hand,  $M_{T2}$ of the background is mainly below the
 top quark mass. Hence, the signal is extracted from the background in
 the high $M_{T2}$ region. 
Since we use the leading jets for $M_{T2}$, there is a combinatorics effect. We found the effect also enhances the signal to background ratio for high $M_{T2}$. 
We perform a detailed simulation with the Matrix Element
 correction to the QCD radiations.  The discovery potential of
the MUED is improved by the $M_{T2}$ cut, and especially, the improvement is significant for the most degenerate
 parameter we consider, $\Lambda R =10$.
 } 
   
\end{center}
\end{titlepage}

\section{Introduction}
In the last decade, extra dimensional models were studied as possible extensions of the Standard Model (SM). The Universal Extra Dimensions model (UED) was developed by Appelquist, Cheng, and Dobrescu \cite{Appelquist:2000nn} (see \cite{Hooper:2007qk} for 
review). 
While other types of extra dimension like the ADD model \cite{ArkaniHamed:1998rs} and
the RS model \cite{Randall:1999ee} are meant to solve the gauge hierarchy problem,
the UED is motivated by the Dark Matter problem
\cite{Servant:2002aq}. One of the most attractive ways to explain Dark
Matter is a new weakly interacting particle (WIMP), and the UED contains WIMP. 

In the UED, all the SM fields propagate in the flat extra dimensions. The fields are expanded in discrete Kaluza-Klein (KK) modes
called KK particles according to the KK number, the extra dimensional
momentum, in the 4D effective theory. In the 5D UED, the extra dimension is compactified on an
$S^1/Z_2$ orbifold to obtain the SM chiral fermions in the zero mode.
This orbifold violates momentum conservation in the extra dimension but
KK parity remains unbroken.  KK number odd (even) modes have odd (even) KK parity,
and the SM particles have the even KK parity. The lightest KK odd particle
(LKP) is stable because this cannot decay into the SM particles due to the
KK parity. 

We consider the Minimal Universal Extra Dimension model (MUED). The mass
spectrum of each KK level at the tree-level is highly degenerate in mass:
mass of the $n$th KK level is approximately $n/R$, where $R$ is the
radius of the compactified extra dimension. The degeneracy is a little
relaxed due to radiative corrections at the one-loop level
\cite{Cheng:2002iz, Georgi:2000ks}. The contribution of the
radiative corrections is large when $\Lambda R$ is large, where
$\Lambda$ is a cutoff scale of the MUED. 
However, $\Lambda R$ cannot be larger than $\Lambda R \sim 40$, because the running gauge coupling of $U(1)_Y$ increases as power law beyond the  MUED scale $1/R$ and blows up at $\sim 40/R$. So the mass spectrum is still nearly degenerate. 

For most choices of parameters, LKP is the first KK photon $\gamma^{(1)}$ which is a good Dark Matter candidate.  
In calculating the relic abundance of LKP $\gamma^{(1)}$ in the nearly degenerate
spectrum, the co-annihilation effect takes an important role \cite{Servant:2002aq, Kong:2005hn,Burnell:2005hm}. The second KK particles also enter in the computation at the one loop level \cite{Kakizaki:2005uy, Kakizaki:2005en, Belanger:2010yx}. The mass scale of LKP Dark Matter consistent with the cosmological observations is $1/R \sim 1.5 \TEV$ \cite{Belanger:2010yx}.

The collider signature of the model with the nearly degenerate spectrum is more
difficult to find because produced visible particles tend to be
soft \cite{Kawagoe:2006sm}. The well-studied new
physics model is Supersymmetry (SUSY) with large mass splittings, and
the decay of the colored SUSY particles produces hard jets and large
$E_T^{miss}$. 
However, in the MUED, soft jets and small $E_T^{miss}$ are mostly generated
due to the spectrum.

The discovery studies at Hadron Colliders have been carried out since
the MUED was proposed
\cite{Cheng:2002ab,Macesanu:2002db,CERN-CMS-CR-2006-062,Bhattacharyya:2009br,Bhatt:2010vm}. 
Previous studies were mainly based on the multilepton channels \cite{Cheng:2002ab,CERN-CMS-CR-2006-062,Bhattacharyya:2009br,Bhatt:2010vm}
and the most promising one is $4l+E_T^{miss}$ in which the background
level is quite low \cite{Cheng:2002ab,CERN-CMS-CR-2006-062,Bhattacharyya:2009br}. 
But the signal events remain very small in this analysis because this
channel is accessible to very limited production processes of MUED.
The analysis including background systematic uncertainties was studied
by the CMS collaboration \cite{CERN-CMS-CR-2006-062}, and the discovery reach
is $1/R \sim 600 \GEV$ with 1 \fb at $\sqrt{s}=14 \TEV$. 
Since the LKP dark matter scenario favors the very high mass scale $m_{LKP} \cong
1/R \sim 1.5 \TEV$ \cite{Belanger:2010yx}, 
it is challenging to look for the signature in $4l+E_T^{miss}$.
In order to check the scenario at the LHC, the sensitivity should be
improved, and therefore, we need an alternative way to search for the MUED.

Multijet channels without requiring multilepton have a statistical
advantage because most MUED events have multiple jets, even though the
analysis based on hard jets and large $E_T^{miss}$ cannot deal with
the signal of the nearly degenerate model due to the softness of jets and small $E_T^{miss}$. The problem of the ordinary
multijet $+ E_T^{miss}$ analysis is that the signal is buried in the SM
background. This is because the top quark pair production $\ttbar$ generates missing particles, neutrinos,  and  hard jets with a large enough cross section.  
If there exists a method that can extract the signal from the background
based on jets, the discovery potential of the MUED could be improved.

We tackle the search in the multijet channel by using a kinematic variable
$\mttwo$ \cite{Lester:1999tx,Barr:2003rg}, sometimes called ``Stransverse
Mass''. $\mttwo$ was originally proposed to measure masses of 
pair-produced particles in the situation with two invisible particles. 
When the true mass of invisible particle is given, $\mttwo$ is bounded by the mass of the produced particles. 
It was proposed that $\mttwo$ can be used not only as a mass measurement
variable but for the event selection \cite{Barr:2009wu, Lester:2007fq},
and this has been applied for the SUSY search by the ATLAS
\cite{daCosta:2011qk} and CMS \cite{CMS-PAS-SUS-11-005} Collaborations.  

In this paper, we point out that $\mttwo$ is effective for the search of
the nearly degenerate model like the MUED. 
To use $\mttwo$ as an event selection, we need to set a test mass for
the invisible particle, and it is set to
zero.  The set test mass is wrong for the MUED events because it is much
smaller than the mass of LKP. This leads to the $\mttwo$
dependence on Up-Stream Radiations (USR). USR is defined as  visible
particles which contribute to the recoil momentum of the subsystem of the pair-produced
particles, and they are mainly initial state radiations (ISR). 
$\mttwo$ of the signal can be large depending on USR, although, without
USR, $\mttwo$ is
small in the nearly degenerate spectrum.
On the other hand, the test mass is correct for the mass of the SM invisible particle,
neutrino. Then, $\mttwo$ of the SM background does not depend on
USR. As shown in Refs.~\cite{Lester:1999tx,Barr:2003rg}, it is mainly below the mass of the
heaviest particle in the SM, the top quark, 
$  \mttwo^{\rm SM} \lesssim m_{top}$.  
Therefore, an excess in the high $\mttwo$ region beyond $m_{top}$ can be seen as the
signal of a nearly degenerate model, and then, $\mttwo$ is effective
to search for the MUED.

In the analysis of this paper, leading two jets in $p_T$ are used to calculate
$\mttwo$. They do not always correspond to jets we want, that
is, we have combinatoric issues when choosing jets for defining
$\mttwo$. Combinatorics smears the $\mttwo$ distribution, and the
smearing effect is different in each process. We found combinatorics
makes $\mttwo$ of the signal larger while $\mttwo$ of the background
does not increase as much as that of the signal. 
Therefore, the smearing effect of combinatorics enhances the signal
excess in the high $\mttwo$ region.

We apply $\mttwo$ to the discovery study of the MUED, and we
require at least one lepton in addition to multijet to avoid the QCD
background. Since the ISR
takes an important role in this method, we perform the event generation
with the Matrix Element correction which evaluates the hard ISR
appropriately. 
This way, we show that the $\mttwo$ analysis improves the
discovery potential compared to the $4l + E_T^{miss}$ analysis.  
\vspace{5pt}

The paper is organized as follows: in section 2 we briefly review the MUED
and describe its LHC signature, the relic abundance of the LKP, and the
experimental constraints. In section 3, we discuss $\mttwo$ for the
event selection in searching for the MUED signal. The simulation setup is
presented in section 4.  We present our analysis result and show that
the discovery potential of the MUED is improved in section 5. Section 6 contains  discussion and conclusion.

\section{The Minimal Universal Extra Dimension model} \label{5dUED}

\subsection{Setup}
 In the case of the 5D UED, there is a 
compactified flat extra dimension in which all the SM
fields universally propagate in addition to the 4D
Minkowski space-time.
Fields are expanded in the
KK modes (KK particles) in the 4D effective theory, and each zero mode
corresponds to the SM particle. 
For example, the 5D real scalar field is decomposed in an infinite
number of the KK modes after integration of the compactified extra dimension $y$,
\begin{eqnarray}
\int d^4x \int^{\pi R}_{-\pi R}&dy& 
\Phi(x,y)\left(\partial^{2}-\partial_{5}^2-m_{SM}^{2} \right)\Phi(x,y)
\\&=&
\int d^4x
 \sum_{n=-\infty}^{\infty}\phi^{(-n)}_{}(x) 
\left(\partial^{2}-\left\{ \left(\frac{n}{R}\right)^{2}+m_{SM}^{2}
 \right \}
 \right)\phi^{(n)}_{}(x)
\\
&=& \int d^4x \sum_{n=-\infty}^{\infty}\phi^{(-n)}_{}(x)
 \left(\partial^{2}-m_{n}^{2} \right)\phi^{(n)}_{}(x)
\end{eqnarray}
where 
\begin{equation}
\Phi(x,y) =\frac{1}{\sqrt{2\pi R}} \sum_{n=-\infty}^{\infty} \phi^{(n)}(x) e^{i\frac{n}{R}y}
\end{equation}
and $m_n^2 = m_{SM}^2 + (n/R)^2$. $R$ denotes the radius of the extra
dimension, and $m_{SM}$ denotes a SM particle mass. The fifth dimensional
momentum is the mass in the 4D effective theory, and
this is much greater than $m_{SM}$, because $1/R \sim O(\rm TeV) $. 
Therefore, we can neglect $m_{SM}$: $m_n \simeq n/R$, which means
the mass spectrum of each KK level is highly degenerate.

Since the simple  compactified
extra dimension $S^1$ gives vector-like fermions, 
an orbifold compactified extra dimension  $S^1/Z_2$ with an
identification of $y \leftrightarrow -y$ is considered in order to obtain chiral fermions
in the zero mode. The orbifold compactification results in another
significant characteristic, the KK parity. KK number is conserved by
virtue of the fifth 
dimensional momentum conservation on $S^1$ compactification, but this is
broken down to the KK
parity  by the orbifold compactification.
The KK parity reflects  ``evenness" and ``oddness" of the KK number. 
All the SM particles have the even KK parity. 
The lightest particle
with the odd KK parity, called the lightest
Kaluza-Klein particle (LKP), is stable since it cannot decay into lighter SM
particles due to its oddness.  The stable LKP, typically the first KK
photon $\gamma^{(1)}$, can be a weakly interacting massive particle
(WIMP) and therefore a good Dark Matter candidate.

To discuss collider phenomenology, we have to determine the mass
spectrum. In this paper, we discuss the Minimal Universal Extra
Dimenison model (MUED). 
The MUED is a minimal extension of
 the 4D SM Lagrangian to the 5D UED. At the cutoff scale $\Lambda$
 it contains only SM fields and no other terms,
 especially no localized terms at two fixed points $y=0, \pi R$ led by orbifold compactification. 
The model parameters of MUED are only three: 5D radius $R$, cutoff scale
 of MUED $\Lambda$, and the SM Higgs
 mass $m_h$.

\subsection{Mass spectrum}
Radiative corrections to masses of the KK modes at the one-loop level were studied in Refs.~\cite{Cheng:2002iz, Georgi:2000ks}.
This correction enlarges mass splitting for each KK level away from the
highly degenerate mass spectrum. The corrected masses are:
\begin{eqnarray}
 m_{X^{(n)}} ^{2}  =\frac{n^{2}}{R^{2}} + m_{X^{(0)}}^{2}+\delta m_{X^{(n)}}^{2} &\mbox{(Boson)},\notag\\
 m_{X^{(n)}} ^{}  =\frac{n^{}}{R^{}} + m_{X^{(0)}}^{} +\delta m_{X^{(n)}}^{}&\mbox{(Fermion)},
\end{eqnarray}
where $m_{X^{(0)}}^{}$ is a SM particle (zero mode) mass. Since $1/R$ is
taken to be larger than $400\GEV$ as mentioned in Sec.~\ref{limit},
$m_{X^{(0)}}^{}$ is much smaller than $1/R$.
The neutral gauge bosons of $U(1)_Y$ and $SU(2)_L$ are 
mixed up in the SM, but mass eigenstates of the KK neutral gauge bosons,
$\gamma^{(n)}$ and $Z^{(n)}$, are nearly $U(1)_Y$ and $SU(2)_L$ 
gauge eigenstates,
$B^{(n)}$ and $W^{3(n)}$, respectively because the diagonal components of mass matrix dominates as  
\begin{equation}
 \begin{pmatrix}
 B^{(n)} 
&
W^{3(n)}
 \end{pmatrix}
 \begin{pmatrix}
 \frac{n^{2}}{R^{2}} + {\delta}m_{B^{(n)}}^{2} + \frac{1}{4}g'^{2}v^{2} & 
 \frac{1}{4}g'gv^{2} 
\\
\frac{1}{4}g'gv^{2}   & \frac{n^{2}}{R^{2}} + {\delta}m_{W^{3(n)}}^{2} + \frac{1}{4}g^{2}v^{2}
 \end{pmatrix}
 \begin{pmatrix}
 B^{(n)} 
\\
W^{3(n)}
 \end{pmatrix}
\end{equation}
where $g'$ is the gauge coupling of $U(1)_Y$, $g$ is that of $SU(2)_L$, and
$v=246 \GEV$ is the vacuum expectation value of the Higgs field. The
radiative corrections to gauge boson masses are given by
\begin{eqnarray}
 \delta m_{B^{(n)}}^{2}&=& -\frac{39}{2} \frac{g'^{2}\zeta(3)}  {16\pi^{2}}
  \frac{1}{R^{2}}+\frac{n^{2}}{R^{2}}  \biggl(-\frac{1}{6} \frac{g'^{2}}{16\pi ^{2}}\biggr) 
 \ln (\Lambda R)^{2} 
\notag\label{142025_13Jun11} 
\\
\delta m_{W^{(n)}}^{2}&=& -\frac{5}{2} \frac{g^{2}\zeta(3)}  {16\pi^{2}}
 \frac{1}{R^{2}}+\frac{n^{2}}{R^{2}}  \biggl(\frac{15}{2} \frac{g^{2}}{16\pi ^{2}}
   \biggr)  \ln (\Lambda R)^{2}
\\
\delta m_{g^{(n)}}^{2} &=& -\frac{3}{2} \frac{g_{s}^{3}\zeta(3)}  {16\pi^{2}} \frac{1}{R^{2}}
+\frac{n^{2}}{R^{2}}  \biggl(\frac{23}{2} \frac{g_{s}^{2}}{16\pi ^{2}}
   \biggr)  \ln (\Lambda R)^{2}\notag\label{141717_13Jun11}
  \end{eqnarray}
  where $\zeta(3)=1.20205 ...$ and $g_s$ is the gauge coupling of $SU(3)_C$.
The second terms in the corrections are dominant, so $m_{W^{(n)}}$ is
lifted, and $m_{B^{(n)}}$ is slightly lowered. 

The mixings of the KK quarks (KK
leptons) are also negligible, 
and they become  $U(1)_Y$ and $SU(2)_L$ 
gauge eigenstates. Neglecting $m_{SM}$, radiative corrections to the KK quarks and  KK leptons are given by
\begin{eqnarray}
 \delta m _{Q^{(n)}} &=& \frac{n}{R} \biggl(3 \frac{g_{s}^{2}}{16\pi ^{2}}
  + \frac{27}{16} \frac{g^{2}}{16\pi ^{2}}  + \frac{1}{16}
  \frac{g'^{2}}{16\pi ^{2}}      \biggr)
  \ln (\Lambda R)^{2}
 \notag\label{141824_13Jun11} 
 \\
 \delta m _{u^{(n)}} &=& \frac{n}{R} \biggl(3 \frac{g_{s}^{2}}{16\pi ^{2}}
  +  \frac{g'^{2}}{16\pi ^{2}}  
  \biggr)
  \ln (\Lambda R)^{2}
  \notag\\
\delta m _{d^{(n)}} &=& \frac{n}{R} \biggl(3 \frac{g_{3}^{2}}{16\pi ^{2}}
  + \frac{1}{4} \frac{g'^{2}}{16\pi ^{2}}  
  \biggr)
  \ln (\Lambda R)^{2}
   \\
  \delta m _{L^{(n)}} &=& \frac{n}{R} \biggl(\frac{27}{16} \frac{g_{}^{2}}{16\pi ^{2}}
  + \frac{9}{16} \frac{g'^{2}}{16\pi ^{2}}  
  \biggr)
  \ln (\Lambda R)^{2}
  \notag \\
 \delta m _{e^{(n)}} &=& \frac{n}{R} \biggl(\frac{9}{4} \frac{g'^{2}}{16\pi ^{2}}\biggr)
  \ln (\Lambda R)^{2}
\notag
\end{eqnarray}
where $Q^{(n)}$ and $L^{(n)}$ denote the $SU(2)_L$ doublet, and $u^{(n)}$, $d^{(n)}$, and $e^{(n)}$ denote
the $SU(2)_L$ singlet.  For the KK top quark we should consider the correction
from its Yukawa coupling, but the production cross section is small. We
do not consider the processes of KK top quark.  

Most KK particles receive positive mass
corrections.  The heaviest particle in each level is $g^{(n)}$   for the
largest correction, and the lightest particle in each level is
typically $\gamma^{(n)}$. Then the LKP is $\gamma^{(1)}$ with the mass $m_{\gamma^{(1)}} \cong 1/R$.  
If the Higgs boson of the SM is as heavy as $m_h \gtrsim 240 \GEV$, the first KK charged Higgs $h^{\pm(1)}$ can be LKP due to the negative
mass correction of the Higgs four point coupling. 
But, of course, this cannot be the Dark Matter.
In this paper, we fix the Higgs  mass at $m_h = 120\GEV$, and we keep
$\gamma^{(1)}$ as LKP so that it is the Dark Matter candidate.

The corrections  are basically proportional to $ \ln \Lambda R$, so the degeneracy is
crucial for the smaller $\Lambda R$.
The cutoff scale of the UED was discussed in \cite{Bhattacharyya:2006ym}, and the
appropriate cutoff scale should be  several dozen $1/R$ for a given
$R$. As the energy scale grows, more KK particles appear, and the
 logarithmic running of the gauge coupling changes into power law running above
 the MUED scale $1/R$. The $U(1)_Y$ gauge 
 coupling blows up (Landau pole) at the energy scale 
 $\sim 40/R$, so we should set the cutoff scale below the Landau pole. 
 The very small $\Lambda R$ is also not appropriate
 because we should consider the higher dimensional operators, and the MUED
 framework is not a good effective theory any more.   
In our analysis, we considered $10 \leq \Lambda R \leq 40$. A benchmark
 point of MUED is chosen as  $1/R=800\GEV$, $\Lambda R =20$, 
and table \ref{benchmark} shows its mass spectrum. 
\begin{table}[htd]
\begin{center}
\begin{tabular}{|c|c|c|c|c|c|c|c|c|c|}
\hline
$m_{\gamma^{(1)}}$ & $m_{ W^{(1)}}$ & $m_{ Z^{(1)}}$& $m_{ e^{(1)}}$ &$m_{ L^{(1)}}$& $m_{ d^{(1)}}$ &$m_{ u^{(1)}}$ &$m_{ Q^{(1)}}$ & $m_{ g^{(1)}}$& \\
800.1 & 847.3 &847.4 & 808.2 & 822.3 & 909.8 & 912.5 & 929.3 & 986.4 & GeV\\
\hline
\end{tabular}
\end{center}
\vspace{-15pt}
\caption{Mass spectrum of first KK level for a benchmark point $(1/R, \Lambda R)=(800,20)$ }
\label{benchmark}
\end{table}%

\subsection{Production and decay at the LHC}

At the LHC, the first KK particles of the odd KK parity are pair-produced,
and  they eventually decay into the LKP.
The dominant production processes are KK gluon+KK quark ($g^{(1)}+Q^{(1)} / q^{(1)}$) 
and KK quark+KK quark $(Q^{(1)}/q^{(1)}+Q^{(1)}/q^{(1)})$. The cross
sections of the colored particles are shown in \cite{Macesanu:2002db}. For
our benchmark point, $\sigma(g^{(1)}+Q^{(1)} / q^{(1)})=12.2$ pb and
$\sigma(Q^{(1)} / q^{(1)}+Q^{(1)} / q^{(1)})=7.4$ pb at $\sqrt{s} =14\TEV$.
The $g^{(1)}$ decays into  $Q^{(1)}Q$ and $q^{(1)}q$ with branching
ratios, BR($g \to Q^{(1)}Q$) $\sim$ 40\% and BR($g \to q^{(1)}q$) $\sim$
60\%, respectively. 
The ratio of inclusive KK quark productions is roughly
$Q^{(1)}Q^{(1)}:q^{(1)}q^{(1)}:Q^{(1)}q^{(1)}=1:1:2$.  
Because $q^{(1)}$ only has the $U(1)_Y$ gauge interaction, 
it directly decays into $\gamma^{(1)}q$. The hard jets mainly come from
this decay. The branching ratios of $Q^{(1)}$ are typically ${\rm
BR}(Q^{(1)}\to Q W^{\pm(1)})\sim 65\%$, ${\rm BR}(Q^{(1)}\to Q
Z^{(1)})\sim 32\%$, and ${\rm BR}(Q^{(1)}\to Q \gamma^{(1)}) \sim 3\%.$
Once $W^{(1)}$ and $Z^{(1)}$ appear from $Q^{(1)}$, they cannot decay
hadronically for kinematical reasons. They democratically decay
into all lepton flavors:
${W^{\pm(1)} \to \gamma^{(1)}l\nu}$ 
and $Z^{(1)}\to \gamma^{(1)}\nu \bar{\nu} \mbox{ or }\gamma^{(1)}l^+ l^-$ through $l^{(1)} \mbox{ or } \nu^{(1)}$.

  This collider signature has been studied in clean channels of
  multilepton such as $4l+ E_{T}^{miss}$
  \cite{Cheng:2002ab,CERN-CMS-CR-2006-062,Bhattacharyya:2009br},
  dilepton, and trilepton  \cite{Bhattacharyya:2009br,Bhatt:2010vm}. The
  missing transverse energy $E_T^{miss}$
is a magnitude of missing transverse momentum $\ptvec{miss}$ which is
measured by the negative sum of transverse momenta of visible particles ,
 $\ptvec{miss} = -\sum \ptvec{jet} -\sum \ptvec{lep}$.
  The leptons arise only from the KK gauge boson $W^{(1)}$ or $Z^{(1)}$ production.
 The $4l + \Etmiss$ channel has been the most 
promising one because the background is extremely small, but the fraction of
  the MUED events going to this channel is about 1\%:
  from the $Q^{(1)}Q^{(1)}$ production, each $Q^{(1)}$ should decay as
  $Q^{(1)}\to Q Z^{(1)} \to Q l^+l^-\gamma^{(1)}$ with the branching
  ratio of 16\%. 

Multijet channels without requiring multileptons are
  statistically advantageous, so we use the multijet + lepton
  channel. This is accessible to about 65\% of the MUED total
  production. The requirement of one lepton is only to avoid the QCD background.

However, we face the difficulty of relativity small
$p_T^{jet}$ and $E_T^{miss}$ due to the small mass splitting between
produced colored particles and LKP.  
Then, the ordinary multijet + $E_T^{miss}$ analysis optimized for
  the typical SUSY expecting hard jets and large $E_T^{miss}$   is not
 effective for the search for the nearly degenerate model: MUED. 
It is important to study a way to squarely address the nearly
  degenerate model using multijet 
because the discovery potential of the MUED could be improved for the statistical advantage. 
In order to search 
 the signal of the nearly degenerate model in the channel we choose, we apply
  $\mttwo$ to an event selection described in Sec.~3.

\subsection{Relic abundance of LKP and experimental constraints }\label{limit}
The LKP Dark Matter relic abundance has been studied in
Refs.~\cite{Servant:2002aq, Kong:2005hn,Burnell:2005hm,Kakizaki:2005uy,
Kakizaki:2005en, Belanger:2010yx}. 
Inclusion of co-annihilation processes is essential for the correct estimation
of the LKP mass. 
In the co-annihilation calculation, the effect of $s$-channel resonances
of the second KK
 particles \cite{Kakizaki:2005uy,Kakizaki:2005en} and the effect of the
 final states with the second KK
particles \cite{Belanger:2010yx} 
 must be considered.
 The MUED mass scale $1/R$ consistent with cosmological observations is
 $1/R \sim 1.5$ TeV including these effects \cite{Belanger:2010yx}. In
 this case,  it is challenging to discover the MUED both at the first
 run ($\sqrt{s}=7 \TEV$) and at the second run ($\sqrt{s}=14 \TEV$) with
 a low luminosity $O$(1) fb$^{-1}$. 
 Hence, it is important to enhance the discovery potential of the MUED
 by developing a new technique and/or a new channel to check the LKP Dark
 Matter scenario at the LHC. 

The electroweak precision test suggests the energy scale of the extra
 dimension $1/R$ should be greater then 550 GeV for $m_h =100$ GeV and
 350 GeV for $m_h =400 \GEV$ at the 95\% C.L. \cite{Gogoladze:2006br,Appelquist:2002wb}. Independent of the Higgs mass, the observed branching ratio of 
$B_d \to X_s \gamma$ constrains  $1/R>$ 600GeV at the 95\%
 C.L. \cite{Agashe:2001xt,Buras:2003mk,Haisch:2007vb}. 

 We have obtained a collider bound on the MUED from the ATLAS SUSY
 search in  multijet +$E_T^{miss}$   with
 35 pb$^{-1}$  \cite{daCosta:2011qk} and 1 \fb
 \cite{Aad:2011ib} data and in multijet + one lepton
 +$E_T^{miss}$ with 1 \fb data \cite{Collaboration:2011iu}.
It gives an upper limit on cross sections after several cuts. We
 have checked which parameters of MUED are excluded using
 Monte Carlo samples generated by \PY \cite{Sjostrand:2006za} \footnote{The event generation of MUED that was used for this bound is described in Sec. 4.1. }, and we found that
$1/R \le 600 \GEV$ with $10 \leq \Lambda \leq 40$ is excluded by the
 multijet +$E_T^{miss}$ analysis \cite{Aad:2011ib} at the 95\%
 C.L.

Then, we focus on $400\GEV \le 1/R \le 1600\GEV$ due to the LKP abundance, 
the branching ratio of $B_d \to X_s \gamma$, the elctroweak precision test,
 and the LHC constraint.

\section{Method for Searching for the MUED}
\subsection{$M_{T2} $ }

Our idea is to apply $M_{T2} $ to the event selection when searching for
the MUED.
The effectiveness of $\mttwo$ to search for SUSY were discussed in \cite{Barr:2009wu, Lester:2007fq}, and $\mttwo$ was
already applied in the search for SUSY in multijet + $E_T^{miss}$ by the
ATLAS collaboration \cite{daCosta:2011qk}. 

We briefly review the definition of $\mttwo$. $\mttwo$, an extension of
transverse mass $M_T$, was originally proposed as a mass measurement
variable in the situation with two invisible particles \cite{Lester:1999tx,Barr:2003rg}. In each event, we only know the total missing
transverse momentum, $\ptmiss$, but each transverse momentum of the
invisible particle cannot be measured.  
The definition of $\mttwo$ is :

\begin{eqnarray}
 \mttwo &\equiv&  \min_{\ptvec{inv(1)} +\ptvec{inv(2)} = \ptvec{miss}}   \left[    \max \left\{ M_T^{(1)}
, \mbox{ } M_T^{(2)}\right\}  \right] \\\notag
\end{eqnarray}
where $M_T$ is defined by 
\begin{eqnarray}
M_T^{(i)}&=&M_T(m_{vis(i)},m_{inv(i)}, \ptvec{vis(1)}, \ptvec{inv(1)} )\notag
\\\\
 &\equiv&\sqrt{ m_{vis(i)}^{2}+m_{inv(i)}^{2}+
2 \left( E_{T}^{vis(i)}E_{T}^{inv(i)} -\ptvec{vis{(i)}} \cdot\ptvec{inv{(i)}}  
						    \right) },\notag
\end{eqnarray}
The transverse energy $E_T^{}$ is given by
\begin{eqnarray}
E_T &\equiv &\sqrt{m^2 + |\ptvec{}|^2}.
\end{eqnarray}
In calculating $\mttwo$, we first construct transverse mass
$M_T^{(i=1,2)}$ and  take the maximum of them for one partition of
$\ptvec{inv{(1)}}$ and $\ptvec{inv{(2)}}$ satisfying $\ptvec{inv{(1)}}+\ptvec{inv{(2)}}=\ptmiss$. 
Then, all the possible partitions are considered, and the minimum value
among them is taken.  

Let us consider the simple case where the same parent
particles are produced and each of them directly decay to a visible
particle and an invisible particle. If the invisible
particle mass $m_{inv}$ is known, $M_T$ is bounded by the parent particle mass, $M_T \leq m_{parent}$ in the correct partition. 
Then, as seen from the definition,
$\mttwo$ is also bounded by the parent particle mass, $\mttwo \leq m_{parent}$. 
We can also interprete $\mttm$ as the invariant mass.     

In practice,  $m_{inv}$ is not known.
In calculating $\mttwo$, we need to set a test mass for the invisible particle.
Many attempts were made
to simultaneously determine the masses of the parent and
invisible particles \cite{Cho:2007qv, Barr:2007hy, Cho:2007dh, Nojiri:2008hy, Burns:2008va, Cohen:2010wv}. One of them \cite{Cohen:2010wv} utilizes the effect of Up-stream
Radiations (USR). 
USR is defined as visible particles which are emitted before parent particles of our interest are produced. 
The transverse momentum of USR, $\Ptvec{}$, is given by,
\begin{eqnarray}
\ptvec{vis(1)} + \ptvec{vis(2)} + \ptvec{miss} = - \Ptvec{} \hspace{5mm}\mbox{(USR)}.
\end{eqnarray}
\begin{figure}
\begin{center}
  \includegraphics[width=70mm]{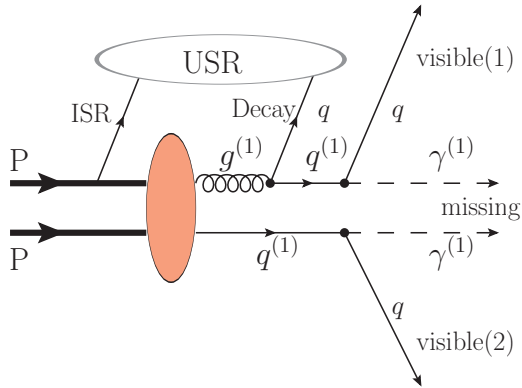}
  \caption{Schematic picture of typical MUED process, $g^{(1)}q^{(1)}$
 production. When two jets from $q^{(1)}\to \gamma^{(1)}q$ and used as
 visible particles to
 construct $\mttwo$, a jet from
 the decay of $g^{(1)}\to q^{(1)}q$ and ISR become USR.}
 \end{center}
\label{USR}
\end{figure}
$\Ptvec{}$ is a measure of the recoil of the parent particles. The
source of USR is mainly initial state radiations (ISR). The decay products can be USR if the decay is before the production of the parent particles.
Fig. 1 illustrates USR in a case of $g^{(1)}q^{(1)}$ production. When we
are interested in $q^{(1)}$, a quark emitted from the decay of $g^{(1)}$
and ISR are considered as USR.

Of course, $\mttm$ has different behaviors depending on whether the test
mass is correct. When we set a correct test mass, $\mttm$
corresponds to the parent particle mass independent of USR. But, when we set a wrong
test mass, $\mttm$ varies with USR. This is because $M_T$ 
 varies with USR and  is no longer 
bounded by the parent particle mass. This property can be used in the search for MUED.

\subsection{$\mttwo$ for the event selection}\label{mt2selection}
The features of $\mttwo$ for the purpose of event selection were studied in
\cite{Barr:2009wu, Lester:2007fq}. We use the $\mttwo$ cut as an event selection
setting the test mass to zero. The set test mass is a correct value for SM because
the only invisible particles of the SM are neutrinos. In this case, $\mttwo$ of
the most background events, especially $\ttbar$ events, is lower than
the top quark mass $\mtop$. This is because we can measure the parent particle mass
with the correct test mass, and the top quark is the heaviest
parent particle in the SM.   
Also, it was found that events without missing momentum
or with fake missing momentum which is parallel to a mismeasured jet have very
small values of $\mttwo$ \cite{Barr:2009wu}.
If a significant excess of  $\mttwo$ above $\mtop$ is observed, it
should be the new physics signal.  

On the other hand, $\mttwo$ of new physics
is not bounded by the parent particle mass because the test mass is
wrong for the Dark Matter candidate of new physics. The upper bound of
$\mttwo$ is a mass combination of the parent
 particle and
the invisible particle in the absence of USR,
\begin{equation}
 \mttm = \frac{m_{parent}^2 -m_{inv}^2}{m_{parent}} \equiv \mu_0.
\label{muzero} 
\end{equation}
In this case, the signal is extracted from the background for models
with a large mass splitting, such as SUSY, but not for models with a degenerate mass spectrum, such as 
$\mu_0 \leq \mtop$, because the signal is buried in the background.

However, considering the recoil momentum of parent particles by USR, 
$\mttwo$ is still a useful variable for the event selection in searching for the nearly
degenerate model: MUED. $\mttm$ varies with USR and can exceed $\mu_0$
due to the wrong test mass.  
For example, when parent particles of same
mass are produced and directly decay to invisible particles emitting
visible particles ($q^{(1)}q^{(1)} \to qq \gamma^{(1)}\gamma^{(1)}$), $\mttm$ 
\cite{Burns:2008va, Cohen:2010wv} is 
\begin{eqnarray}
 \mttm &=& \sqrt{\mu(P_T)^2+P_T \mu(P_T)}\ge \mu_0 ,\label{mttmmu}
\\\notag\\
\mbox{where }\hspace{5mm}\notag&
\\ 
\mu(P_T)&\equiv&
 \mu_0\left(\sqrt{1+\left(\frac{P_T}{2m_{parent}}\right)^2}
  -\frac{P_T}{2m_{parent}}\right)
\\
&\to& \mu_0 \left(\frac{m_{parent}}{P_T}\right)\hspace{1cm}\mbox{for
 }P_T\gg m_{parent} \notag
\label{endpoint2}
\end{eqnarray}
where  $P_T$ is the magnitude of the momentum of USR.
There is a rich source of USR because processes of heavy
particles tend to come along with hard QCD radiations including ISR.
Hard ISR gives large recoil of produced particles, and $\mttm$ can have
a large value depending on USR. Note that the background events do not have
$\mttwo$ dependence on USR because the test mass is correct, so most events are kept lower than $\mtop$.

When analyzing events, we cannot tell the origins of visible particles:
whether the particles come from decays of heavier particles or are
QCD radiations. Practically, leading two jets in $p_T$ are
used as visible particles to construct $\mttwo$. If they
correspond to two ``correct'' particles, namely if each particle is a decay
 product of each pair-produced particle, $\mttwo$ behaves as discussed
above. However, the leading particles can be decay products of one parent
particle, and also hard ISR can be one or both of the leading particles. These
cases are called ``combinatorics''. 

In many events, $\mttwo$ of the leading particles corresponds to
$\mttwo$ of the correct particles. For instance, the rate of correspondence is about half for  $q^{(1)}q^{(1)}$ or $\ttbar$ as
shown later. 
Combinatorics smears the $\mttwo$ distribution. The
smearing effect is significant for high $\mttwo$, and it is different in each
process.

\begin{figure}[h!]
\begin{center}
\includegraphics[width=80mm]{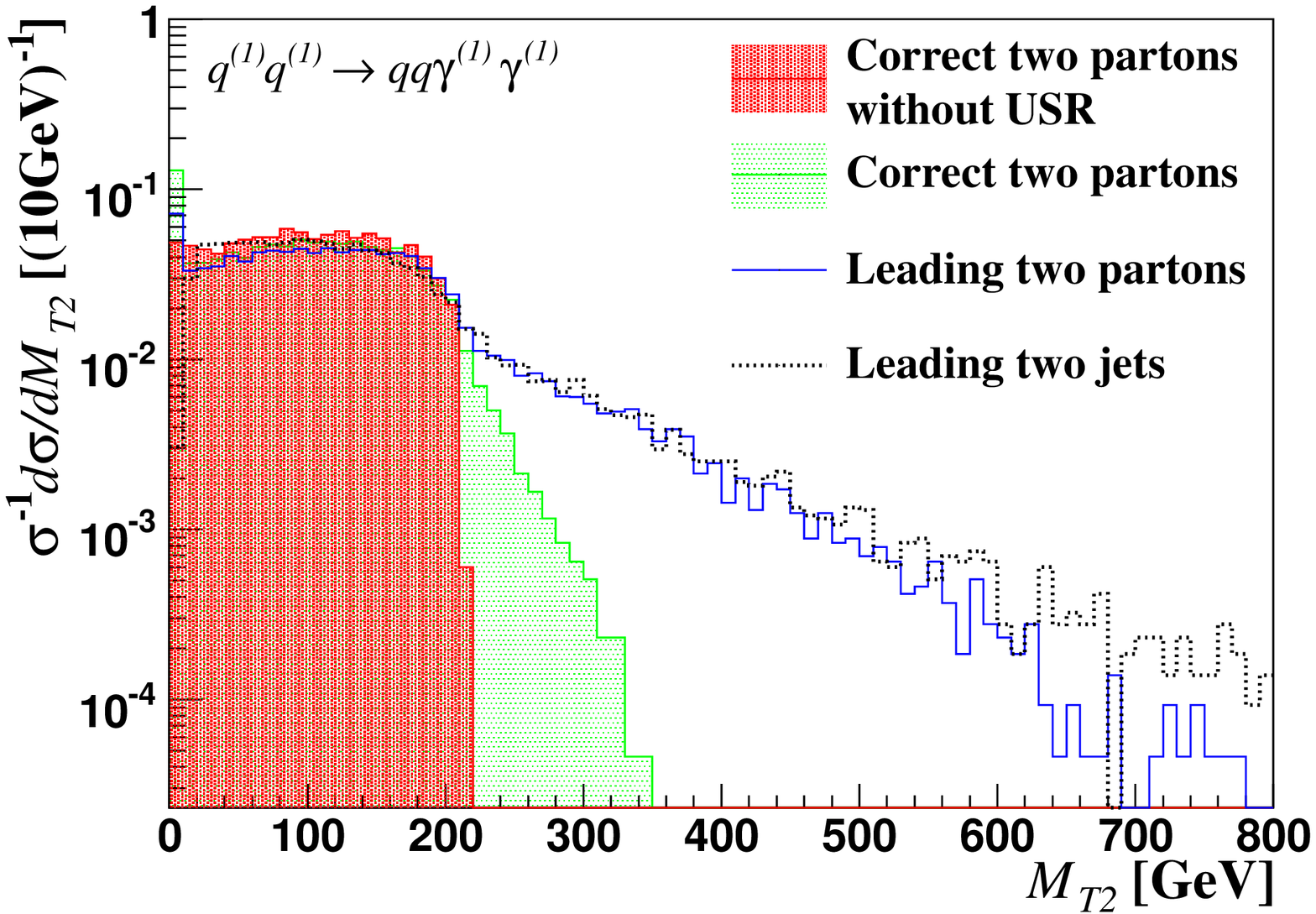}
\includegraphics[width=80mm]{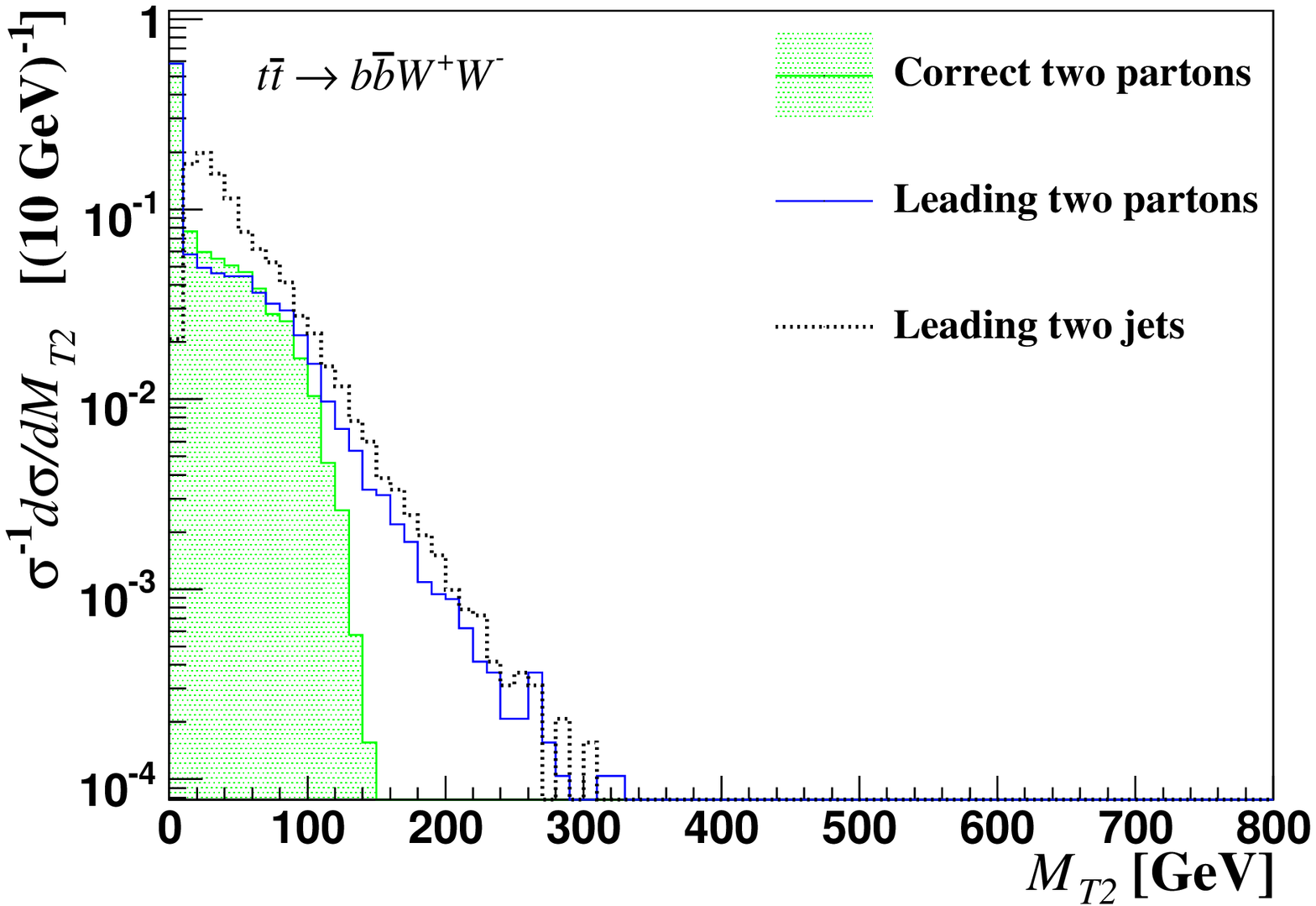}
\label{mt2evsl }
\caption{Distributions of $\mttwo$ for $q^{(1)}q^{(1)}\to qq
 \gamma^{(1)} \gamma^{(1)}$ in the left and $\ttbar \to \bbbar W^+W^-$
 in the right generated by \MG \cite{Alwall:2007st} where $m_{q^{(1)}} = 912.5$
GeV, $m_{\gamma^{(1)}} = 800.1$GeV, and $\mu_0 = 211.0$GeV. $\mttwo$
 is constructed with the correct two
 partons (green shaded area), the leading two partons (blue solid line), and
 leading two jets (dotted line).
 Also, events  were generated without additional jets, that is, without
 USR in the parton level, and $\mttwo$ was calculated with the correct two partons (red shaded area). The distributions are normalized to 1.}
\end{center}\label{2figmt2}
\end{figure}

In order to see the effects of USR and combinatorics, we generated 
the $q^{(1)}q^{(1)}$ production of the MUED benchmark point  and the $\ttbar$
production adding up to one jet in the parton level \footnote{The events were
generated by \MG at $\sqrt{s}=14$ TeV. The fragmentation and
hadronization were simulated by {\sc Pythia 6.4},
and the detector effects were simulated by PGS 4 \cite{PGS}. The simulation setup is
the same as the simulation described in Sec. 4 except that
here only one jet was added as the Matrix Element correction. The MLM
matching \cite{Alwall:2007fs} was prescribed.}, and
constructed $\mttwo$ with the correct partons and with the leading partons/jets. 
These $\mttwo$ distributions are shown in fig. 2.

The green 
(red) shaded area shows $\mttwo$ with (without) an additional jet in
the parton level using correct two partons: quarks from
the direct decay $q^{(1)} \to q \gamma^{(1)}$ for $q^{(1)} q^{(1)}$  and $b$
quarks from $t \to bW $ for $\ttbar$. LKPs and neutrinos produce $\ptvec{miss}$. 
For the signal, $\mttwo$ without USR (red shaded area) is bounded by
the mass combination $\mu_0 =210$ GeV given by (\ref{muzero}).
Including USR (green shaded area) $\mttwo$ varies with $P_T$ and
exceeds $\mu_0$ as (\ref{mttmmu}), and theoretically it reachs about 440
GeV with extremely large $P_T$. 
For the $\ttbar$ background, $\mttwo$ of correct two partons with USR (green
shaded rarea) does not exceed $\mtop$ as expected. 

\begin{table}[htd]
\begin{center}
\begin{tabular}{|c|c|c|}
\hline
 Parton level & $q^{(1)}q^{(1)}\to qq
 \gamma^{(1)} \gamma^{(1)}$ + 0, 1 jet & $\ttbar \to \bbbar W^+W^-$ + 0,
 1 jet \\
\hline
 $\mttwo^{leading} =\mttwo^{correct}$& 61.6\% &49.1\%  \\
\hline
$\mttwo^{leading} >\mttwo^{correct}$ & 30.3\% &22.4\%  \\
\hline
$\mttwo^{leading} <\mttwo^{correct}$ & 8.1\% &28.5\%  \\
\hline
\end{tabular}
\end{center}
\vspace{-15pt}
\caption{The evaluation of combinatrics for $q^{(1)}q^{(1)}$ of the
 benchmark point and $\ttbar$. $\mttwo$ is constructed in the parton level. We compare $\mttwo$ of leading partons and correct partons in each event. }
\label{combi}
\end{table}%

When the leading partons are used for $\mttwo$ (blue solid line),
combinatorics smears the $\mttwo$
distribution. $\mttwo$ of the leading partons spreads over the endpoint of $\mttwo$ of correct partons
to roughly double the value at the endpoint as in fig. 2. Table \ref{combi} shows the evaluation of
combinatorics. The leading partons correspond to the correct partons for
60\%  of the time for
$q^{(1)}q^{(1)}$ and for 50\% for $\ttbar$. The smearing effect due to
combinatorics is different in each process: $\mttwo^{leading} >\mttwo^{correct}$ for three quarters of
the combinatorics events of $q^{(1)}q^{(1)}$, while $\mttwo^{leading} <\mttwo^{correct}$ for more
than half of the combinatorics events of $\ttbar$. Therefore, combinatorics
assists to enhance the signal to background ratio for high $\mttwo$. 

Also, the detector effects are simulated after the fragmentation and
hadronization, and $\mttwo$ is constructed with the leading two jets
(dotted line). The distribution is similar with the $\mttwo$
distribution with the leading two partons.  

The dependence on USR makes the signal excess in the high $\mttwo$ region
even for the nearly degenerate model, and the smearing effect of combinatorics enhances the
excess. 
It can been seen that events with high $\mttwo$, say above 200 GeV, are
dominated by the $q^{(1)}q^{(1)}$ signal over the $\ttbar$ background. Since for the other
SM background processes the parent particle is lighter than the top quark,
those background events are expeced to have $\mttwo$ lower than
$\mtop$. 
Hence, $\mttwo$ is an effective event selection to search for the nearly
degenerate model.

\section{Simulation}
 \subsection{Signal}

 Monte Carlo (MC) samples of MUED signal were generated both by
a private implementation in \MG (MG/ME) \cite{Alwall:2007st} and an implementation
\cite{ElKacimi:2009zj} in \PY{ }6.4   \cite{Sjostrand:2006za}. CTEQ5.1L
was used as the leading-order (LO)  parton distribution function
(PDF). In the case of MG/ME, the Matrix Element was calculated by HELAS
\cite{Murayama:1992gi}, and the fragmentation and hadronization were simulated with \PY. 
 The effects of jet reconstruction and detector smearing were simulated
 through PGS 4 \cite{PGS}. 

  We consider $1/R$ from 400 GeV to 1.6 TeV in steps of 100 GeV with
 $\Lambda R$ = 10, 20, 30, and 40.
 The remaining parameter, $m_h$, is set to 120 GeV. 
  The MUED spectrum is simplified by neglecting $m_{SM}$ for the first KK level.
  The processes we consider are pair productions of the colored first KK particles, $g^{(1)}$, $q^{(1)}$, and $Q^{(1)}$. The signal events 
 corresponding to  luminosities of 5 fb$^{-1}$ at $\sqrt{s}$ = 7
 TeV and of more than 10 fb$^{-1}$ at $\sqrt{s}$ =14 TeV were generated
 by \PY. Table \ref{ProdProc} shows the production processes which can be generated
 in the \PY{ }implementation.  
 
  \begin{table}[h]\centering
\begin{center}
\begin{tabular}{| l | l |l|}
\hline
\multicolumn{2}{|c|}{ Process} & flavor\\
\hline
 KK gluon + KK gluon	&$ {g + g \rightarrow g^{(1)} + g^{(1)}}$ &\\
 \hline
 KK quark + KK gluon	&$ {g + q \rightarrow g^{(1)} + Q^{(1)};\  g^{(1)} + q^{(1)}}$&\\
\hline
 KK quark + KK quark&$ {q_i + q_j \rightarrow Q^{(1)}_{i} + Q^{(1)}_{j};\  q^{(1)}_{i}  + q^{(1)}_{j}}$& all $i,j$\\
& $ {q_i + q_j \rightarrow Q^{(1)}_{i} + q^{(1)}_{j}}$ &   all $i,j$\\
\hline
KK quark + KK antiquark	 &$ {g + g \rightarrow Q^{(1)} + \bar Q^{(1)};\ q^{(1)} + \bar q^{(1)}}$& \\
 &$ {q + \bar q \rightarrow Q^{(1)} + \bar Q^{(1)};\ q^{(1)} +\bar q^{(1)}}$ & \\
& ${q_i + \bar q_j \rightarrow Q^{(1)}_{i} + \bar q^{(1)}_{j}}$&  $i \ne j$\\
 &$ {q_i + \bar q_j \rightarrow Q^{(1)}_{i} + \bar Q^{(1)}_{j};\ q^{(1)}_{i} + \bar q^{(1)}_{j}}$ & $i \ne j$ \\
& $ {q_i + \bar q_i \rightarrow Q^{(1)}_{j} + \bar Q^{(1)}_{j}}$&  $i \ne j$\\
\hline
\end{tabular}
\caption{Processes of MUED generated by the \PY{ }implementation  \cite{ElKacimi:2009zj}. $SU(2)_L$ doublet and singlet 
first KK quark are denoted by $Q^{(1)}$ and $q^{(1)}$ respectively.}
\label{ProdProc}
\end{center}
\end{table}
 
  Since we use the $\mttwo$ dependence on USR, the ISR has an important
  role. In order to reliably evaluate the hard ISR, we considered the Matrix Element correction in MG/ME adding up to one jet to the pair productions. The MLM
  matching \cite{Alwall:2007fs} was applied to remove the overlap between jets from the Matrix Element and ones from the Parton Shower. 
This prescription was demonstrated for the benchmark point of $\Lambda
  R=20$ and  $1/R =800\GEV$. The spectrum of this point is listed in table \ref{benchmark}. 
However it is very time consuming to generate all of the signal MC
  samples with the Matrix Element correction, so we used \PY{ }rather than MG/ME to generate
  them for the discovery study. 
We will show in Sec.~\ref{EVSelection}. that MC samples generated by ME/ME with
  the Matrix Element correction have larger excess over background than ones generated by
  \PY{ }for the benchmark point. Hence, the event generation by \PY{ }is conservative.

\subsection{Background}
MC samples of the SM background, $\ttbar$, $W/Z+jets$, Diboson ($WW, WZ, \mbox{and } ZZ$), etc., were produced with 
MG/ME  using the PDF set CTEQ6.1L, and fragmentation and
hadronization were simulated with \PY{ }in the same way of the signal. 
For $\ttbar$, $W/Z+jets$, and Diboson, up to two partons were added in
the Matrix Element and the MLM matching was prescribed.
The MC samples were detector-simulated through PGS 4. 
The dominant background processes, $\ttbar$ and $W/Z+jets$, were
normalized to the next-leading-order (NLO) cross section consistent
with the inclusive dijet analysis of the ATLAS MC study \cite{Aad:2009wy}. 

For the sake of comparison with the $4l + E_T^{miss}$ analysis, we generated some multilepton background processes, such as four leptons through off-shell $Z^{\ast}$ or $\gamma^{\ast}$. 
The luminosities of generated SM background are more than 2 \fb  at
$\sqrt{s}$= 7 and more than 10 \fb at $\sqrt{s}$= 14 TeV, respectively. 
The summary of the background processes is shown in table \ref{SMBG}.

\begin{table}[ht!]
\centering
\begin{tabular}{|l |r| r|r|r|}
\hline
 & \multicolumn{1}{| c|}{$\sqrt{s}=7\TEV$ } &\multicolumn{1}{| c|}{ $\sqrt{s}=14\TEV$ }\\
\cline{2-3}
{ Process } 
& {  {\boldmath $\sigma$}$\times$  efficiency}  
& {  {\boldmath $\sigma$}$\times$  efficiency}  \\
\hline \hline
    $\ttbar + 0, 1, 2$ jets &   130 pb  
	 & 765 pb   \\    
     $(W \to l\nu)  + 1, 2$ jets$^{\dagger}$& 287 pb &  897 pb \\   
     $(Z \to l^{+} l^{-},\nu\bar{\nu})  + 1, 2$ jets$^{\dagger}$& 98.8 pb &  341 pb \\    
    $W^+ W^- + 0, 1, 2$ jets 	&    35.0 pb 	
	 &  97.5 pb  \\
    $WZ + 0, 1, 2$ jets 	&    15.0 pb 	 &  44.1 pb  \\
 $ZZ$ +0, 1, 2 jets 	& 4.78 pb&  12.9 pb \\
 $Z^{\ast}/\gamma^{\ast} Z^{\ast}/\gamma^{\ast} \to 2l^+2l^-$ & 29.2 fb
     &   57.2 fb \\
$Z+\bbbar$ &35.5 pb    & 135 pb  \\
$W+\bbbar$ &22.5 pb & 55.0 pb \\
$(Z/\gamma^{\ast} \to l^+l^-,\nu\bar{\nu})+\ttbar$ & 28.7 fb &205 fb\\
$(W \to l\nu) +\ttbar$ & 37.2 fb  & 131 fb  \\
\hline
\end{tabular}
\caption{Summary of the SM datasets used in this analysis. +n jets are
 added in Matrix Elements with the MLM matching. 
$^{\dagger}$MC samples of $W/Z + jets$ are preselected as $p_T^{\rm 1st jet} >90 \GEV$.
The cross sections listed are calculated with the preselected MC samples
 generated by {\sc MadGraph/MadEvent 4.4}.  
}
\label{SMBG}
\end{table}

\section{Analysis}
\subsection{Object selection}
The object selection is that an electron and a muon are required to have
$p_T>10 \GEV$ and $|\eta|< 2.5$ and a jet is required to have $p_T>20\GEV$ and $|\eta| <2.5$. 
In order to avoid recognizing a shower from an electron as a jet, a jet within
$\Delta R <0.2$ ($\Delta R = \sqrt{\Delta \eta^2 + \Delta \phi^2}$) from any electron is removed.
Charged leptons from hadronic activity also should be removed. If an electron and a jet are found within $0.2 <\Delta R < 0.4$, the jet is kept
and the electron is rejected 
Similarly, if a muon and a jet are found within $\Delta R <0.4$, the muon is rejected.

\subsection{Event selection} \label{EVSelection}

First, we require two jets with $p_T^{jet} >$ \{100, 20 GeV\}, and
$p_T^{miss}$ ($E_T^{miss}$) 
must exceed 100 GeV. Cuts stronger than these will reduce the MUED signal.
At least one charged lepton, an electron or a muon, with $p_T^{lep} > 20 \GEV$ is required in
our analysis.
The cuts imposed above, particularly the requirement of one lepton,  are necessary to avoid the QCD background. Either
$\{p_T^{jet1}>100\mbox{ GeV}, p_T^{miss}>100\mbox{ GeV}\}$ or
$p_T^{lep}>20$ GeV is used as a trigger.
When there is only one lepton, 
we impose a cut of $M_T^{lep,miss}>100$ GeV to reduce the $W+jets$ background, where
\begin{eqnarray}
M_T^{lep,miss} \equiv \sqrt{2 \left( p_T^{lep}p_T^{miss}- \ptvec{lep}\cdot \ptvec{miss} \right)}\hspace{5mm}. \notag
\end{eqnarray}
Finally, we construct $\mttwo$ with the leading two jets, and impose $\mttwo >200 \mbox{ GeV}$.
To summarize our event selection: 
\begin{itemize}
  \item CUT1: $p_T^{jet} > \{100, 20 \mbox{ GeV}\}$
  \item CUT2: $E_T^{miss}>100 \mbox{ GeV}$
  \item CUT3: At least one lepton with $p_T^{lep} > 20 \mbox{ GeV}$ 
  \item CUT4: If the number of lepton is one, $M_T^{lep,miss}>100 \mbox{ GeV}$ 
  \item CUT5: $\mttwo >200 \mbox{ GeV}$. 
\end{itemize}

In order to demonstrate the effectiveness of $\mttwo$ for the MUED, we only
use the basic cuts 1-4 except one on $\mttwo$. The cuts 1-4 are comparable with ones imposed in the ATLAS and CMS new physics searches in  one lepton + jets + $E_T^{miss}$ with low luminosity at $\sqrt{s}=7 \TEV$ \cite{Bernet:2011rd,Aad:2011hh}.
We do not use the  $M_{eff}$
cut and the $E_T^{miss}/M_{eff}$ cut which are used to extract the
signal especially by the ATLAS collaboration in the search for SUSY \cite{Aad:2009wy},
where 
\begin{equation}
M_{eff} \equiv \sum^4_{jet}p_T+\sum _{lepton}p_T+ E_T^{miss}.
\end{equation}
It is common that the $\Delta \phi_{jet,miss}$ cut is applied to reduce
events with fake missing due to the mismeasurement of jets, but this is 
not necessary because the later $\mttwo$ cut has a similar role \cite{Barr:2009wu}. 

\begin{table}[htdp!]
\begin{center}
\begin{tabular}{|c|c|r r r r r| }
\hline
\multicolumn{2}{|c|}{Process} & CUT1 & CUT2& CUT3 & CUT4 & CUT5 {\footnotesize (Optimal)} \\
\hline
 $g^{(1)}+g^{(1)}$ 
  	& MG/ME &1,028 & 832 & 119 & 62& 25\\
\cline{2-2}
	& PYTHIA & 937 &757 &108 &63 &22 \\
\hline
 $g^{(1)}+q^{(1)}/Q^{(1)}$ 
 & MG/ME &9,196 & 7,218& 1,234& 675&  241\\
\cline{2-2}
& PYTHIA & 8,569 & 6,694 &1,344 & 731 &223\\
\hline
 $q^{(1)}/Q^{(1)}$ 
 & MG/ME&5,315 &4,035  &863 &508 &148\\ 
\cline{2-2}
$+q^{(1)}/Q^{(1)}$
& PYTHIA & 4,497 & 3,276  &690 	&436 &84\\
\hline
 $q^{(1)}/Q^{(1)}$ 
 & MG/ME  &1,444 &1,075  &206  &115  &27 \\
\cline{2-2}
$+\bar{q}^{(1)}/\bar{Q}^{(1)}$
& PYTHIA & 1,301 &955 & 163	&112 &20\\
\hline
Total MUED
& MG/ME & 16,983 & 13,160 & 2,422 & 1,360& 441\\
\cline{2-7}
& PYTHIA & 15,304 & 11,682 & 2,305 &1,342 &349 \\
\hline \hline
\multicolumn{2}{|c|}{$\ttbar$ } & 426,074 & 57,533 &23,239	&5,620 & 243\\
\multicolumn{2}{|c|}{$W$ } 	&400,527 & 97,907 &35,386	&1,031 & 85\\
\multicolumn{2}{|c|}{$Z$ } 		&142,368 & 53,801 &916		&107	&12\\
\multicolumn{2}{|c|}{$W/Z+\ttbar /\bbbar$ }& 1,121 & 304 &103 & 49	&10\\
\multicolumn{2}{|c|}{Diboson } &29,141 &4,482 & 1,335	&252	&40\\
\hline
\multicolumn{2}{|c|}{Total Standard Model} &999,231 &214,027 &60,979 &7,059 &390\\
\hline \hline
Total MUED& MG/ME & 0.05 &   0.17 & 0.06& 0.78	&4.10\\
\cline{2-7}
 $Z_B$& PYTHIA & 0.05 & 0.14 & 0.05 &0.77 	& 3.37 (7.57) \\
\hline 
\end{tabular}
\end{center}
\vspace{-15pt}
\caption{Cut flow for 1 \fb at $\sqrt{s}$ = 14TeV. The MUED  benchmark point
 is $\{1/R, \Lambda R \}=\{ 800\mbox{ GeV},20 \}$. The MUED signal generated by
 \MG(MG/ME) with the MLM matching is normalized to one generated by \PY. 
$\mttwo >$ 350 GeV is an optimal cut that maximizes the
 significance $Z_B = 7.57$.  }
\label{800mass}
\end{table}

The cut flow in table \ref{800mass} shows that the $\mttwo$ cut (CUT5) significantly
reduces the SM background to a level comparable to the MUED. Since the Matrix Element correction increases event rates in the high $\mttwo$ region, there
remain more signal events generated by MG/ME after CUT5 than ones
generated by \PY. Therefore, the signal rate in the fast event generation by \PY{ }is 
a little underestimated and hence is conservative.  

\begin{figure}[h]
\begin{center}
\includegraphics[width=80mm]{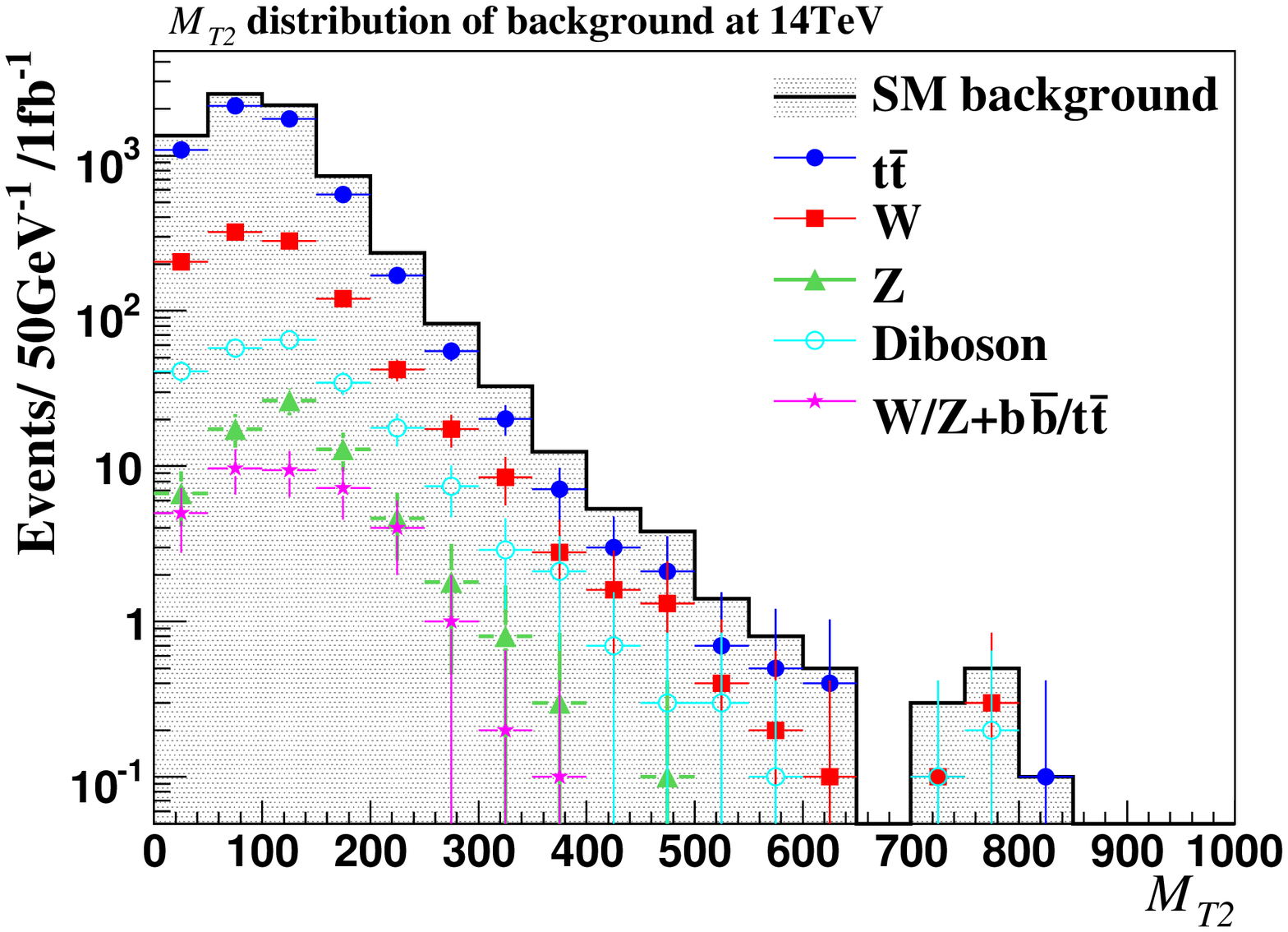}
\includegraphics[width=80mm]{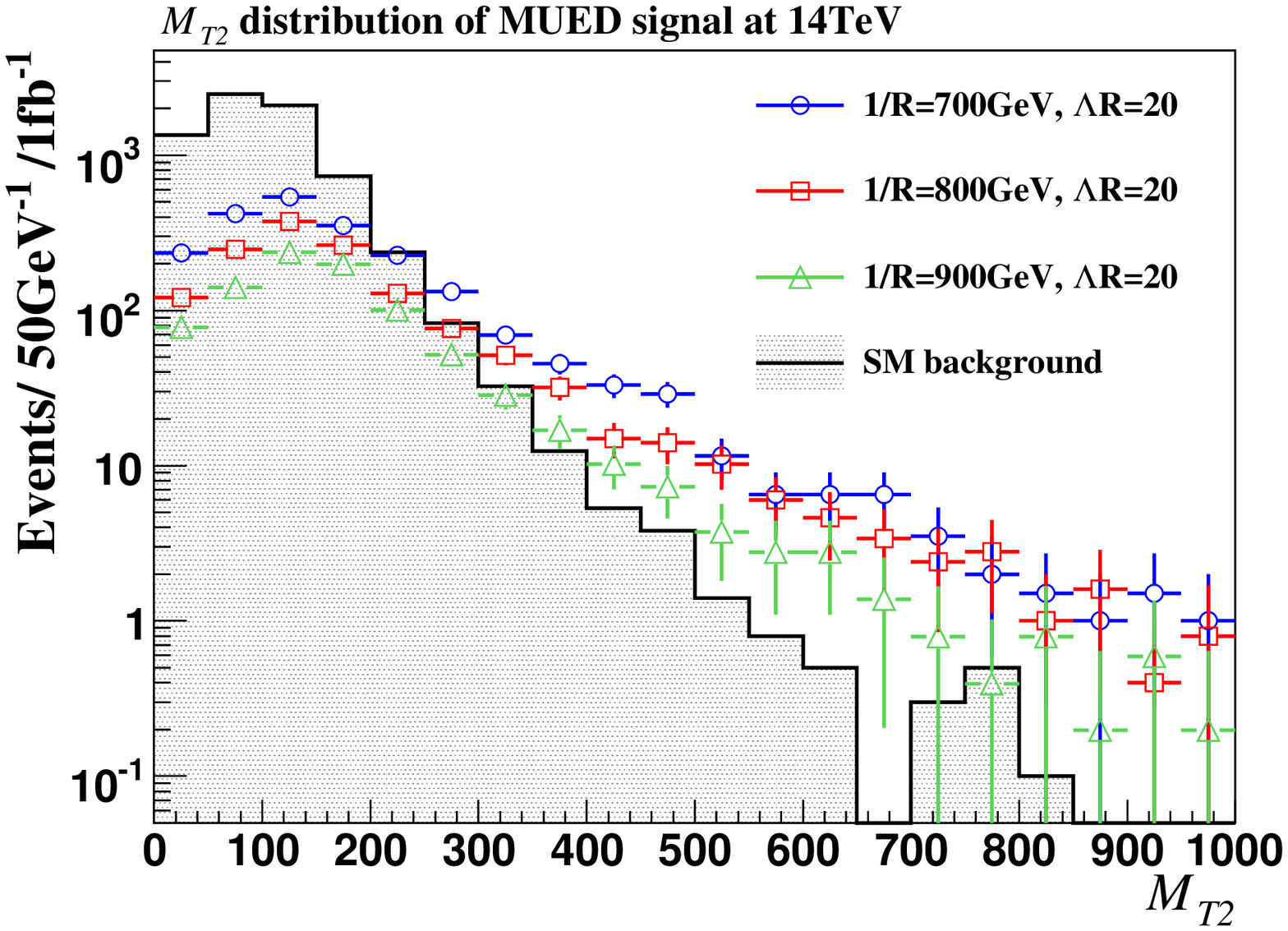}
\caption{Distributions of $\mttwo$ of the leading two jets after CUT4 for each SM background in the left, and
 total SM background and MUED signal points of $1/R= 700, 800, 900 \GEV$
 and $\Lambda R =20$ generated by \PY{ }in the right.  }
\label{BG2}
\end{center}
\end{figure}

Fig. \ref{BG2} shows that the dominant background are $\ttbar$,
$W+jets$, and Diboson. 
Background events that remain even after CUT5  mainly come from combinatorics. 
The peak of MUED events is $\mttwo<200\GEV$, but the signal events have
a long
tail which can be understood as a result of  the variant endpoint due to
the wrong test mass discussed in section \ref{mt2selection}. {There is
combinatorics for both the signal and the background, and especially it tends to increase $\mttwo$ of the signal events. Combinatorics help to
enhance the signal excess.}
As a result, we can successfully extract the signal from the
background even based on jets.

\subsection{Discovery potential}
For the study of the discovery potential it is necessary to take
systematic uncertainties into account in addition to statistical uncertainties. 
We use the significance $Z_B$ \cite{Linnemann:2003vw}, which is provided
by the {ROOT} library \cite{Brun:1997pa}, using the same approach as in
the ATLAS discovery study of the SUSY \cite{Aad:2009wy}. $Z_B$ is calculated using a convolution of a Poisson and a Gaussian term to account for systematic errors. 
For the backgrounds except those from QCD, a reasonable estimate of the systematic
uncertainty is $\pm 20\%$. 
The discovery potential is studied by finding the optimal $\mttwo$ cut (in step of 50GeV) to maximize the significance $Z_B$.
We define ``discovery''
 when $Z_B> 5$ and more than 10 signal events remain after the cuts.

In order to compare the $\mttwo$ analysis in the multijet + lepton mode with the previously studied
$4l+E_T^{miss}$ analysis, we also check the discovery potential in $4l+E_T^{miss}$ 
using the same MC samples and using the same definition of discovery. In
the $4l+E_T^{miss}$ analysis, the following cuts are imposed \cite{Cheng:2002ab}:(1) four isolated leptons with $p_T^{lep} > \{35, 20, 15, 10 \GEV\}$ 
are required, (2) $E_T^{miss} > 50 \GEV$, and
(3) an invariant mass $M_{ll}$ for all possible pairs of opposite
sign same flavor leptons and remove events if $|M_{ll}-m_Z| <10 \GEV$ to reduce
background from the $Z$ boson. The estimated background from our MC samples is
10 events/100 fb$^{-1}$. The fake leptons should be considered to
evaluate the background level of $4l+E_T^{miss}$ more appropriately, but
the fake leptons are not considered since they are not important for our analysis based on jets.

\begin{figure}[h]
\begin{center}
\includegraphics[width=120mm]{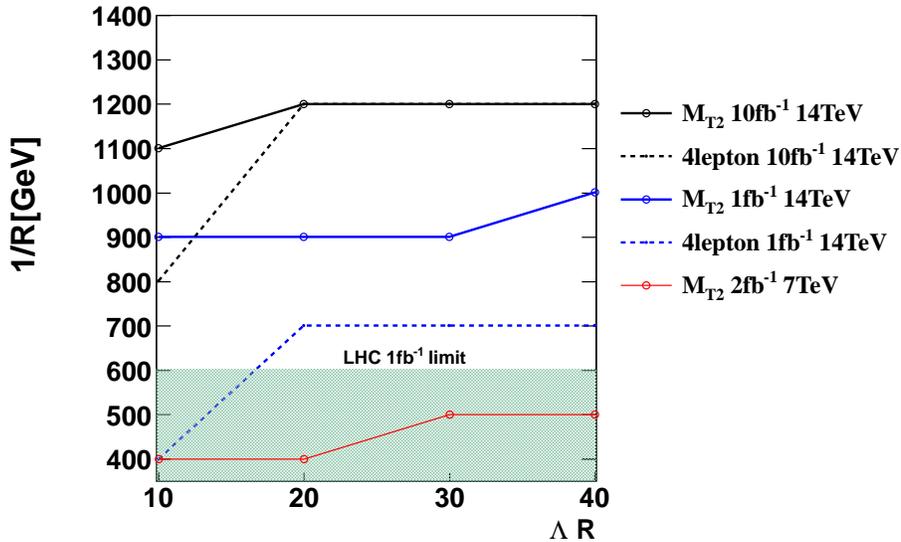}
\caption{Discovery potential of the MUED with 1 \fb and 10 \fb at
$\sqrt{s}$ = 14 TeV in the $4l+E_T^{miss}$ analysis and the $\mttwo$ analysis and discovery potential of  2 \fb at  $\sqrt{s}$ = 7 TeV
only in the $\mttwo$ analysis. 
  For a given luminosity, the parameter region below the line will be discovered.}
\label{RlambR}
\end{center}
\end{figure}

The spectrum is more degenerate for smaller $\Lambda R$, which is more difficult for discovery in general. 
Note that for fixed $1/R$, the MUED with smaller $\Lambda R$ has a larger
cross section simply because  the KK gluon and the KK quark become lighter as in
Eqs.~(\ref{142025_13Jun11}) and (\ref{141824_13Jun11}).
Fig.~4 shows that the discovery potential does not vary with
changing $\Lambda R$ in
the $\mttwo$ analysis.  
The first run of LHC at $\sqrt{s}=7\TEV$ will have an integrated luminosity of 2 \fb and
so it can discover up to $1/R \sim 500 \GEV$. However this parameter
region was already excluded by the ATLAS multijet$ + E_T^{miss}$ analysis
with 1 fb$^{-1}$, as mentioned in Sec.~\ref{limit}. 
 The second run at 14 TeV will discover up to $1/R \sim 1\TEV$ with 1 \fb
 and $1/R \sim 1.2 \TEV$ with 10 fb$^{-1}$. 

In the $4l+E_T^{miss}$ analysis, the discovery reach at 14 TeV is $1/R \sim
 700\GEV$
 with 1 \fb and $1/R \sim 1.2 \TEV$ with 10 \fb  for $20 \le \Lambda R \le
 40$, but the sensitivity is very low for $\Lambda R =10$ : the discovery
 reach is only $1/R = 400\GEV$ with 1 \fb and $1/R = 800\GEV$ with 10 fb$^{-1}$.

The result shows that our $\mttwo$ analysis improves the discovery
 potential. In particular, the improvement is so significant for the
 most degenerate parameter $\Lambda R = 10$  that the discovery
 potential improves from $1/R=400\GEV$ to 900 GeV.

\section{Discussion and Conclusion}

We have improved the discovery potential of the MUED. We include the background 
systematic uncertainties. The previously studied multilepton channels, such as
$4l+E_T^{miss}$, are clean channels but only sensitive to the limited
number of processes. 
On the other hand, we utilize the multijet + lepton channel,
and it is accessible to the most processes of the MUED.
Although this channel is statistically advantageous, it  difficult to
extract the MUED signal events from backgrounds. 
However, we succeeded in  extracting  the signal from
the background by applying $\mttwo$ for the event selection. This is
because the signal events can have large values of $\mttwo$ thanks for the dependence on USR with the wrong test mass, and because the combinatorics effect enhances the signal excess. 

The simulation result indicates that the discovery potential is significantly
 improved by using $\mttwo$ cut. The most important achievement  is that our $\mttwo$ analysis
has a much greater sensitivity for the MUED for the most degenerate mass
 spectrum than the $4l + E_T^{miss}$ analysis.
For 1 \fb at $\sqrt{s} = 14\TEV$, the discovery potential increases 
by 500 GeV in terms of $1/R$ for the fixed $\Lambda R = 10$.
Therefore, the $\mttwo$ analysis is particularly effective for the case
 of a highly degenerate spectrum.

The relic abundance of the Dark Matter implies that the very high value of
$1/R \sim 1.5 \TEV$ is favored, while our analysis reaches only up to $1/R\sim 1.2 $ TeV with 10 fb$^{-1}$ at $\sqrt{s} = 14 \TEV$. 
However, there is possibility to improve the discovery potential further. 
We considered the Matrix Element correction and the NLO correction to the background. If those corrections to the signal are included, the Matrix Element correction enhances the event rate for high $\mttwo$ as shown in Sec.~\ref{EVSelection}, and the NLO correction increases the cross section. 
Also, additional cuts, such as the $b$-jet veto and $E_T^{miss}/M_{eff}$, could enlarge
the discovery potential of the MUED, although we did not introduce them in
order to emphasize the
effectiveness of $\mttwo$.  The $b$-jet veto must be particularly
effective because it significantly reduces the $\ttbar$ background. 
Including these effects the discovery reach will be improved, and the Dark Matter favored MUED may be searched with $O(100)$ fb$^{-1}$. 

The search using $\mttwo$ has a model independent aspect.
The benefit of the $\mttwo$ analysis is significant when there is large USR. 
We might expect
large USR because 
a common feature of many new
physics models is production of heavy colored particles associated with hard ISR contained in USR.
This analysis  
is also applicable to other models:  SUSY with R parity or Little Higgs model with T
parity with a nearly degenerate mass spectrum.

For our analysis, we required one lepton to avoid the QCD background, but this is not always necessary. If the well-understood QCD MC samples are prepared, we can study the 
discovery potential based on the $\mttwo$ analysis in the inclusive dijet channel, which is statistically advantageous and is accessible to
models that rarely emit leptons. 

 Because of the significant improvement of the search for MUED by using
 $\mttwo$, this work implies that the LHC will be able to have better sensitivities to nearly degenerate models.
A more general analysis with a nearly degenerate spectrum is left to a future work.

\section*{Acknowledgement}
We thank Shoji Asai, Shigeki Matsumoto, Seong Chang Park, Ryosuke Sato and Satoshi Shirai for useful discussions. 
This work is supported by  the World Premier International Research Center Initiative (WPI initiative) MEXT, Japan.  
HM was supported in part by the U.S. DOE under Contract DE-AC03-76SF00098, in part by the NSF under grant PHY-04-57315, and in part by the Grant in-Aid for scientic research (23540289) from Japan Society for Promotion of Science (JSPS). MN was also supported by the Grant-in-Aid for scientific research
(22540300) from JSPS.
\bibliographystyle{kuma}
\bibliography{reference}

\end{document}